\documentclass[letterpaper,12pt]{article}
\usepackage{tabularx} 
\usepackage{amsmath,amssymb}  
\usepackage{graphicx} 
\usepackage[margin=1in,letterpaper]{geometry} 
\usepackage{cite} 
\usepackage[final]{hyperref} 
\hypersetup{
	colorlinks=true,       
	linkcolor=blue,        
	citecolor=blue,        
	filecolor=magenta,     
	urlcolor=blue         
}
\usepackage{blindtext}
\usepackage{graphicx}
\graphicspath{ {./figures/} }
\usepackage{float}
\usepackage{tensor}

\begin{document}

\begin{titlepage}

\begin{center}
\hfill TU-1217\\
\hfill KEK-QUP-0036
\vskip 1.in

\renewcommand{\thefootnote}{\fnsymbol{footnote}}

{\Large \bf
Probing Gauss-Bonnet-Corrected Inflation with \\ \vspace{2.5mm} 
Gravitational Waves
}

\vskip .5in

{\large
Kamil Mudruňka$^{(a)}$\footnote{mudrukam@fjfi.cvut.cz}
and
Kazunori Nakayama$^{(b,c)}$\footnote{kazunori.nakayama.d3@tohoku.ac.jp}
}

\vskip 0.5in

$^{(a)}${\em 
FNSPE, Czech Technical University in Prague, Břehová 7, 115 19 Praha 1, Czech Republic
}

\vskip 0.2in

$^{(b)}${\em 
Department of Physics, Tohoku University, Sendai 980-8578, Japan
}

\vskip 0.2in

$^{(c)}${\em 
International Center for Quantum-field Measurement Systems for Studies of the Universe and Particles (QUP), KEK, 1-1 Oho, Tsukuba, Ibaraki 305-0801, Japan
}

\end{center}
\vskip .5in

\begin{abstract}

The low energy effective action of quantum gravity may include the higher curvature terms such as the Gauss-Bonnet term.
The inflaton dynamics may be affected by the Gauss-Bonnet term if there is an inflaton-Gauss-Bonnet coupling.
We show that an inflation model with a simple power law potential is made viable if it is coupled to the Gauss-Bonnet term since the prediction on the scalar spectral index and the tensor-to-scalar ratio are modified.
We further point out that such a model predicts huge amount of gravitational waves at the high frequency range around 100\,GHz--100\,THz through the perturbative inflaton decay into gravitons induced by the Gauss-Bonnet term.
Thus the spectrum of high frequency gravitational background is a unique feature of the inflation models with a Gauss-Bonnet correction.

\end{abstract}

\end{titlepage}

\tableofcontents

\renewcommand{\thefootnote}{\arabic{footnote}}
\setcounter{footnote}{0}

\section{Introduction}

Out of the many cosmological inflation models in existence, many have already been ruled out by the constraints from the cosmic microwave background (CMB) observation by the Planck experiment~\cite{Planck:2018jri}. This includes the simplest scalar inflaton model with a quadratic potential~\cite{Linde:1983gd}.\footnote{
	See, however, Refs.~\cite{Destri:2007pv,Nakayama:2013jka,Nakayama:2013txa,Nakayama:2014wpa} for polynomial modification of the inflaton potential in order to make the inflation model viable. 
}
However, this might not be the case if besides pure General Relativity (GR), there are other higher order terms present in the inflaton plus gravitational field action. In this article we study extension of GR by the Gauss-Bonnet term coupled to the inflaton field. Such extension may come from the string theory~\cite{Antoniadis:1993jc}, and hence the introduction of Gauss-Bonnet term is theoretically well motivated.

There are many studies on the cosmology with the Gauss-Bonnet term~\cite{Kawai:1998bn,Hwang:1999gf,Nojiri:2005vv,Nojiri:2005jg,Carter:2005fu,Satoh:2007gn,Kawai:1998ab,Satoh:2008ck,Guo:2009uk,Guo:2010jr,Jiang:2013gza,Kawai:2017kqt,Chakraborty:2018scm,Odintsov:2018zhw,Odintsov:2019clh,Pozdeeva:2020apf,Pozdeeva:2021iwc,Kawai:2021bye,Kawai:2021edk,Khan:2022odn,Tsujikawa:2022aar,Odintsov:2023aaw,Odintsov:2023weg,Kawai:2023nqs}.
Among various possible effects of the Gauss-Bonnet term, we focus on its role on the correction to the slow-roll inflation dynamics.
It is known that, if there is an inflaton coupling to the Gauss-Bonnet term, the predictions of the density perturbation of the Universe, i.e., the scalar spectral index and the tensor-to-scalar ratio and so on, are modified~\cite{Satoh:2007gn,Guo:2010jr,Jiang:2013gza,Chakraborty:2018scm,Pozdeeva:2020apf,Pozdeeva:2021iwc} compared with the case of standard potential-driven inflation.

The purpose of this paper is twofold. 
First, we show that the inflation with a simple monomial power-law potential, such as just a quadratic one, can become consistent with the CMB observation once the correction from the Gauss-Bonnet term is taken into account.
Second, such an inflaton-Gauss-Bonnet coupling, necessary to make the inflation model viable, induces the decay of the inflaton into the graviton pair in the reheating stage~\cite{Ema:2021fdz}. 
It makes a huge gravitational wave (GW) background in the present universe at the high-frequency range around 1\,meV--1\,eV scale.
It is a very unique prediction of this particular model, which makes it possible to observationally distinguish this model from the most other inflation models.

This paper is organized as follows. 
In Sec.~\ref{sec:inflation} we briefly review the inflaton dynamics and summarize the formulae for evaluating the curvature perturbation in the presence of Gauss-Bonnet coupling.
In Sec.~\ref{sec:gw} we calculate the characteristic GW spectrum predicted in the model through the perturbative inflaton decay and show that the high frequency GW amplitude is so huge that it is can be a smoking gun  signature of the model.  
Sec.~\ref{sec:conc} is devoted to conclusions and discussion.
For readers' convenience, comparisons of the formulae for the curvature perturbation with other literature are summarized in Appendix~\ref{sec:comparison}. 

Throughout this paper we use the following convention. The derivative with respect to the conformal time $a\left(t\right)d\tau=dt$ will be denoted by $\prime$ with $a(t)$ being the cosmic scale factor, while the dot will be kept for the derivative with the cosmological time, e.g., $a' = a \dot a$.

\section{Inflation with Gauss-Bonnet term} \label{sec:inflation}

In this section we review the inflation with Gauss-Bonnet term correction. 
First we derive the background dynamics of the inflaton and then study the scalar and tensor perturbations generated during inflation. In the calculation of the power spectra we follow the procedure described in Ref.~\cite{Satoh:2008ck}.

\subsection{Inflaton dynamics}

The model we study in this article is described by the following action
\begin{gather*}
   S=\int d^4 x \sqrt{-g}\left[\frac{1}{2}M_{\rm pl}^2R - \frac{1}{2}\partial^\mu \phi \partial_\mu \phi-V\left(\phi\right)-\frac{1}{16}\xi\left(\phi\right)R_{\rm GB}^2\right],
\end{gather*}
where $R$ is the  Ricci curvature, $M_{\rm pl}$ the reduced Planck scale, $\phi$ is the inflaton field, $V(\phi)$ the inflaton potential, $\xi(\phi)$ is some function of the inflaton which is taken arbitrary in this section and $R^2_{\rm GB}$ denotes the Gauss-Bonnet term,
\begin{gather*}
   R_{\rm GB}^2=\tensor{R}{^\mu^\nu^\alpha^\beta}\tensor{R}{_\mu_\nu_\alpha_\beta}-4\tensor{R}{^\mu^\nu}\tensor{R}{_\mu_\nu}+R^2.
\end{gather*}
Due to the inflaton field coupling to the Gauss-Bonnet term, it is no longer a topological term and contributes to the field equations in a non-trivial way. The field equation is \cite{Nojiri:2005jg}
\begin{equation}
\label{eqTotalGravFieldEq}
M_{\rm pl}^2\tensor{G}{_\mu_\nu}+\frac{1}{4}R\nabla_\mu \nabla_\nu\xi+\frac{1}{2}\tensor{G}{_\mu_\nu}\nabla^2\xi-\left(\nabla^{\mathstrut}_\rho\nabla^{\mathstrut}_{(\mu}\xi\right)\tensor{R}{_{\nu)}^\rho}+\left(\frac{1}{2}\tensor{g}{_\mu_\nu}\tensor{R}{^\rho^\sigma}-
\frac{1}{2}\tensor{R}{_\mu^\rho_\nu^\sigma}\right)\nabla_\rho\nabla_\sigma\xi=\tensor{T}{_\mu_\nu},
\end{equation}
where $G_{\mu\nu}$ is the Einstein tensor and $T_{\mu\nu}$ is the energy-momentum tensor.
For now we consider arbitrary coupling function $\xi\left(\phi\right)$ and inflaton potential $V\left(\phi\right)$. Later we will restrict our calculation to a specific choice. For the flat Friedmann-Lemaitre-Robertson-Walker (FLRW) universe the modified Friedmann equation is
\begin{equation}
\label{eqDerivedFriedmann1}
3 M_{\rm pl}^2 \mathcal{H}^2-\frac{3}{2}\frac{\mathcal{H}^3\xi^{\prime}}{a^2}=\frac{1}{2}{\phi^{\prime}}^2+a^2 V\left(\phi\right),
\end{equation}
where $\mathcal{H}=\frac{a^{\prime}}{a}$, or alternatively in terms of cosmological time
\begin{equation}
\label{eqFriedmannMetric}
3 M_{\rm pl}^2 H^2-\frac{3}{2}H^3\dot{\xi}=\frac{1}{2}\dot{\phi}^2+V\left(\phi\right),
\end{equation}
where $H=\frac{\dot a}{a}$ is the Hubble parameter.
The evolution equation for $\phi$ can be easily obtained from the action with the FLRW metric. Using the fact that $\ddot{a}=a\left(\dot{H}+H^2\right)$, the Euler-Lagrange equation for this action is given by
\begin{equation}
\label{eqFriedmannField}
\ddot{\phi}+3H\dot{\phi}+\frac{\partial V}{\partial\phi}+\frac{3}{2}\frac{\partial\xi}{\partial\phi}H^2\left(\dot{H}+H^2\right)=0.
\end{equation}


We postulate that the slow-roll inflation happens if the following conditions are satisfied: 
\begin{equation}
\label{eqSlowRollCondition}
\left|\frac{\ddot{\phi}}{H\dot{\phi}}\right|\ll 1,\quad\frac{\dot{\phi}^2}{M_{\rm pl}^2 H^2}\ll 1,\quad \left|\frac{H\dot{\xi}}{M_{\rm pl}^2}\right|\ll 1,\quad \left|\frac{\dot{H}}{H^2}\right|\ll 1.
\end{equation}
Under the slow-roll conditions, we obtain the following slow-roll equations modified by the Gauss-Bonnet term coupling as
\begin{equation}
\label{eqSlowRollFriedmann}
3 M_{\rm pl}^2H^2=V,\quad \dot{\phi}=\frac{1}{3H}\frac{\partial V}{\partial \phi}-\frac{1}{2}H^3\frac{\partial \xi}{\partial \phi}.
\end{equation}
Following Ref.~\cite{Satoh:2008ck}, we define the slow-roll parameters as follows and demand that absolute values of them remain very small during the inflation,
\begin{align}
\epsilon=\frac{M_{\rm pl}^2}{2 V^2}\left(\frac{\partial V}{\partial\phi}\right)^2,
\quad
\eta=\frac{M_{\rm pl}^2}{V}\frac{\partial^2 V}{\partial\phi^2},
\quad
\alpha=\frac{1}{4 M_{\rm pl}^2}\frac{\partial V}{\partial\phi}\frac{\partial \xi}{\partial\phi},
\quad
\beta=\frac{V}{6 M_{\rm pl}^2}\frac{\partial^2 \xi}{\partial\phi^2},
\quad
\gamma=\frac{V^2}{18 M_{\rm pl}^6}\left(\frac{\partial \xi}{\partial\phi}\right)^2.
\label{epsilon}
\end{align}
Note that $\gamma = 4\alpha^2/(9\epsilon)$.


\subsection{Scalar perturbations}

We calculate the perturbations of \eqref{eqTotalGravFieldEq} in the gauge $\delta\phi=0$, i.e., the inflaton field looks spatially homogeneous. The general scalar perturbation around a flat FLRW metric in conformal time can be decomposed as
\begin{equation}
ds^2=a\left(\tau\right)^2\left[-\left(1+2A\right)d\tau^2+2\partial_i B dx^i d\tau + \left(\delta_{ij}+2\psi\delta_{ij}+2\partial_i\partial_j E\right)dx^i dx^j\right].
\end{equation}
In this decomposition we consider the perturbation generated by $E$ to be traceless i.e. $\nabla^2 E = \partial_1^2 E + \partial_2^2 E + \partial_3^2 E = 0$. To the first order in the perturbations, the Einstein equations are
\begin{gather}
\begin{split}
2\left(1-\frac{\sigma}{2}\right)\nabla^{2}\psi-6\mathcal{H}\left(1-\frac{3}{4}\sigma\right)\psi^{\prime}-2\mathcal{H}\left(1-\frac{3}{4}\sigma\right)\nabla^2 B=2a^2AV-3\mathcal{H}^2\sigma A,
\end{split}\\
\begin{aligned}
-2\mathcal{H}\left(1-\frac{3}{4}\sigma\right)A-2\left(1-\frac{\sigma}{2}\right)\psi^{\prime}=0,
\end{aligned}\\
\begin{aligned}
\left(1-\frac{\sigma}{2}\right)A-\left(1-\frac{\ddot{\xi}}{2 M_{\rm pl}^2}\right)\psi+\frac{1}{a^2}\left[\left(1-\frac{\sigma}{2}\right)a^2B\right]^{\prime}-\frac{1}{a^2}\left[\left(1-\frac{\sigma}{2}\right)a^2E^{\prime}\right]^{\prime}=0.
\end{aligned}
\end{gather}
We already took $\delta\phi=0$, but we can impose one more gauge condition, since two gauge conditions can be imposed on the scalar perturbations.
By further imposing the gauge condition $E=0$, a single equation for $\psi$ can be extracted
\begin{gather}
\label{eqScalarPertPsiOnly}
A_\psi^2\psi^{\prime\prime}+\left(A_\psi^2\right)^{\prime}\psi^{\prime}-C_\psi^2 A_\psi^2\nabla^2\psi=0,
\end{gather}
where
\begin{align}
&C_\psi^2=\frac{2}{A^2_\psi}\left[\left(\frac{a^2 X^2}{\mathcal{H}Y}\right)^{\prime}-a^2\left(1-\frac{\ddot{\xi}}{2}\right)\right], \label{Cpsi}\\
&A^2_\psi=6a^2 X\left[1-\left(1-\frac{\rho^2}{6}-\sigma\right)\frac{X}{Y^2}\right], \label{Apsi} \\
&\sigma\equiv\frac{H\dot\xi}{M_{\rm pl}^2},\quad \rho^2\equiv\frac{\dot{\phi}^2}{M_{\rm pl}^2H^2}\mathrm{,}\quad X\equiv1-\frac{\sigma}{2}\mathrm{,}\quad Y\equiv1-\frac{3}{4}\sigma.
\end{align}
Note that, under the slow-roll approximation, we find
\begin{align}
	\sigma = -\frac{4}{3}\alpha-\gamma,~~~~~~\rho^2=2\epsilon + \frac{4}{3}\alpha + \frac{1}{2}\gamma.
\end{align}
After the Fourier expansion of $\psi$ as
\begin{align}
	\psi=\int\frac{d^3 k}{\left(2\pi\right)^3}e^{ik\cdot x} \psi_k,
\end{align}
it yields the following Mukhanov equation for the canonical variable $\Psi_k\equiv M_{\rm pl}A_\psi \psi_k$:
\begin{equation}
\label{eqScalarPertPsiCanonical}
\Psi_k^{\prime\prime}+\left(C^2_\psi k^2-\frac{A_\psi^{\prime\prime}}{A_\psi}\right)\Psi_k=0.
\end{equation}

The canonical quadratic action for the scalar perturbation is given by
\begin{align}
	S &= \int d\tau\int\frac{d^3k}{(2\pi)^3} \frac{A_\psi^2 M_{\rm pl}^2}{2}\left[\left|\psi'_{k}\right|^2 - C_\psi^2 k^2 \left|\psi_{k}\right|^2 \right]\\ 
	&=\int d\tau\int\frac{d^3k}{(2\pi)^3} \frac{1}{2}\left[\left|\Psi'_{k}\right|^2 - \left(C_\psi^2 k^2-\frac{A_\psi''}{A_\psi}\right) \left|\Psi_{k}\right|^2 \right].
\end{align}
Note that we need $A_\psi^2 > 0$ and $C_\psi^2 > 0$ to avoid the ghost and gradient instability. 
In the model we consider in this paper, these conditions are satisfied. In order to derive the power spectrum, we need to fix the normalization condition, which requires quantum field treatment. We define the creation and annihilation operator as
\begin{align}
	\Psi_k (\tau) = \widetilde\Psi_k (\tau) a_{\vec k} +  \widetilde\Psi^*_k (\tau) a^\dagger_{-\vec k},
\end{align}
where $\left[a_{\vec k} ,  a^\dagger_{\vec k'} \right] = (2\pi)^3\delta(\vec k-\vec k')$. Requiring the Bunch-Davies boundary condition, the solution to the equation of motion of $\widetilde\Psi_k(\tau)$ is given by
\begin{align}
	\widetilde\Psi_k (\tau) = e^{\frac{i(2\nu_\psi+1)\pi}{4}} \frac{1}{\sqrt{2C_\psi k}} \sqrt{\frac{-\pi C_\psi k\tau}{2}} H_{\nu_\psi}^{(1)}(-C_\psi k\tau),
\end{align}
under the slow roll approximation that the derivative of $C_\psi$ is negligible, where
\begin{align}
	\nu_\psi^2 = \frac{1}{4} + \frac{\tau^2 A_\psi''}{A_\psi} \simeq \frac{9}{4} + 9\epsilon -3\eta- \alpha -3\beta,
\end{align}
and hence $\nu_\psi \simeq 3/2 + 3\epsilon-\eta -\alpha/3-\beta$.
Thus the long wavelength limit of $\Psi_k$ should look like $\Psi_k(\tau) \to (-C_\psi k\tau)^{-\nu_\psi}$.
The overall normalization of the power spectrum is determined with the standard procedure for deriving the power spectrum of the curvature fluctuation with quantum field treatments, 
\begin{align}
	\mathcal P_\psi (k) = \frac{\mathcal P_\Psi (k)}{A_\psi^2 M_{\rm pl}^2} 
	\simeq \frac{1}{2\epsilon+\frac{4\alpha}{3}+\frac{\gamma}{2}} \left(\frac{H_{\rm inf}}{2\pi M_{\rm pl}}\right)^2 \left(-k\tau\right)^{3-2\nu_\psi},
\end{align}
where $H_{\rm inf}$ denotes the Hubble scale around which the perturbation with cosmological scales exit the horizon. Thus the scalar spectral index $n_s-1$ for this model is
\begin{equation}
	n_s-1=3-2\nu_\psi= -6\epsilon+2\eta+\frac{2}{3}\alpha+2\beta.  \label{ns}
\end{equation}
The terms with $\alpha$ and $\beta$ represent corrections to the standard slow-roll inflation prediction due to the Gauss-Bonnet term. 
According to the result of the Planck observation, we need~\cite{Planck:2018jri}
\begin{align}
	\mathcal P_\psi (k_0) = \frac{1}{2\epsilon+\frac{4\alpha}{3}+\frac{\gamma}{2}} \left(\frac{H_{\rm inf}}{2\pi M_{\rm pl}}\right)^2 \simeq 2.1\times 10^{-9},
    \label{Ppsi}
\end{align}
and the scalar spectral index $n_s$ should be around $0.96$--$0.97$. 
Detailed analysis will be done in the next section.

\subsection{Tensor perturbations}

Next we consider the tensor perturbation of the form
\begin{equation}
\label{eqTensorPertMetric}
ds^2=a\left(\tau\right)^2\left[-d\tau^2+\left(\delta_{ij}+h_{ij} \right)dx^i dx^j\right],
\end{equation}
where $h_{ij}$ obeys the transverse-traceless condition $h^i_i=\partial_i h_{ij}=0$. The resulting first order equation reads
\begin{equation}
\label{eqTensorPert}
\left(1-\frac{\sigma}{2}\right){h}_{ij}^{\prime\prime}+\left[2\mathcal{H}-\frac{1}{2}\frac{\left(\mathcal{H}\xi^{\prime}\right)^{\prime}}{a^2 M_{\rm pl}^2}\right]{h}_{ij}^{\prime}-\left(1+\frac{\sigma}{2}-\frac{1}{2}\frac{\xi^{\prime\prime}}{a^2 M_{\rm pl}^2}\right)\nabla^2 h_{ij} = 0.
\end{equation}
In order to quantize the perturbations we expand them into
\begin{equation}
h_{ij}=\sum_{\lambda\in\left\lbrace +,\times\right\rbrace}\int\frac{d^3 k}{\left(2\pi\right)^3}e^{ik\cdot x}h_{k,\lambda} \epsilon^\lambda_{ij},
\end{equation}
where the sum runs over two GW polarizations which we denote by $+$ and $\times$ and $ \epsilon^\lambda_{ij}$ denotes the polarization tensor that satisfies $\epsilon_{ij}^\lambda \epsilon_{ij}^{\lambda'} = \delta_{\lambda\lambda'}$.

The quadratic action for the graviton is given by
\begin{align}
	S &=  \sum_\lambda\int d\tau\int\frac{d^3k}{(2\pi)^3} \frac{A_T^2 M_{\rm pl}^2}{8}\left[\left|h'_{k,\lambda}\right|^2 - C_T^2 k^2 \left|h_{k,\lambda}\right|^2 \right]\\ 
	&= \sum_\lambda\int d\tau\int\frac{d^3k}{(2\pi)^3} \frac{1}{2}\left[\left|h^{c\,\prime}_{k,\lambda}\right|^2 - \left(C_T^2 k^2-\frac{A_T''}{A_T}\right) \left| h^c_{k,\lambda}\right|^2 \right],
\end{align}
where we have defined the canonical graviton $h^c_{k,\lambda}\equiv M_{\rm pl}A_T h_{k,\lambda}/2$ and 
\begin{align}
	&A_T^2\equiv a^2\left(1-\frac{\sigma}{2}\right),  \label{AT}\\
	&C_T^2\equiv \frac{a^2}{A_T^2}\left(1+\frac{\sigma}{2}-\frac{\xi^{\prime\prime}}{2a^2 M_{\rm pl}^2}\right)  \label{CT}.
\end{align}
Note that we need $A_T^2 > 0$ and $C_T^2 > 0$ to avoid the ghost and gradient instability.\footnote{
	The so-called Gauss-Bonnet inflation without an inflaton potential~\cite{Kanti:2015pda} is excluded by this condition~\cite{Hikmawan:2015rze}. See also Ref.~\cite{Tsujikawa:2022aar} for analysis with more general nonlinear Gauss-Bonnet terms in the action.
}
In the model we consider in this paper, these conditions are satisfied.
The equation of motion is given by
\begin{gather}
{ h^{c\,\prime\prime}_{k,\lambda}}+\left(C^2_T k^2-\frac{A_T^{\prime\prime}}{A_T}\right)h^c_{k,\lambda}=0.
\end{gather}
Similarly to the case of scalar perturbation, the tensor perturbation is expressed in terms of the creation and annihilation operator as
\begin{align}
	h^c_{k,\lambda} (\tau) = \widetilde h_{k,\lambda} (\tau) a_{\vec k,\lambda} +  \widetilde h^*_{k,\lambda} (\tau) a^\dagger_{-\vec k,\lambda},
\end{align}
where $\left[a_{\vec k,\lambda} ,  a^\dagger_{\vec k',\lambda'} \right] = (2\pi)^3\delta(\vec k-\vec k')\delta_{\lambda\lambda'}$. 
Thus the solution to the equation of motion of $\tilde h_k(\tau)$ is given by
\begin{align}
	\widetilde h_{k,\lambda} (\tau) = e^{\frac{i(2\nu_T+1)\pi}{4}} \frac{1}{\sqrt{2C_T k}} \sqrt{\frac{-\pi C_T k\tau}{2}} H_{\nu_T}^{(1)}(-C_T k\tau),
\end{align}
where
\begin{align}
	\nu_T^2 = \frac{1}{4} + \frac{\tau^2 A_T''}{A_T} \simeq \frac{9}{4} + 3\epsilon + \alpha.
\end{align}
Thus we have $\nu_T \simeq 3/2 + \epsilon + \alpha/3$.
Repeating the same procedure as the scalar case, we obtain the power spectrum of the tensor perturbation as
\begin{align}
	\mathcal P_h(k) = \frac{4\mathcal P_{\tilde h}(k) }{A_T^2 M_{\rm pl}^2} \simeq  
	8\left(\frac{H_{\rm inf}}{2\pi M_{\rm pl}}\right)^2 \left(-k\tau\right)^{3-2\nu_T}.
\end{align}

Under the slow roll approximation, the tensor spectral index $n_T$ and tensor-to-scalar ratio $r$ for this model is
\begin{align}
&n_T=3-2\nu_T=-2\epsilon-\frac{2}{3}\alpha,\\ 
&r\equiv \frac{\mathcal P_h}{\mathcal P_\psi} \simeq 16\epsilon+\frac{32}{3}\alpha+4\gamma.   \label{nT}
\end{align}
So far there is no severe observational constraint on the tensor spectral index $n_T$. 
On the other hand, the tensor-to-scalar ratio should be smaller than about $0.05$~\cite{Planck:2018jri}. Detailed analysis will be done in the next section.

\section{Gravitational wave background}
\label{sec:gw}

In this section we show that the inflation model with a simple power law potential can be consistent with the CMB observation after the Gauss-Bonnet correction is taken into account. 
Then we show that the Gauss-Bonnet term leads to the inflaton decay into the gravitons during the reheating phase, so that the very abundant high frequency GW background is the unique prediction of this model.
To be concrete, we restrict ourselves to the following choice of $\xi\left(\phi\right)$ and $V\left(\phi\right)$:
\begin{equation}
\xi\left(\phi\right)=8\xi_0 e^{-\lambda\phi/M_{\rm pl}},~~~~~~ V\left(\phi\right)=\frac{1}{2}m^2\phi^2.
\end{equation}
Thus we have three model parameters: $\xi_0, \lambda$ and $m$.
In Sec.~\ref{sec:viable}, these parameters are determined so that the model is consistent with the CMB observation.
In Sec.~\ref{sec:high}, we compute the high frequency graviton spectrum produced by the inflaton decay by using the parameters determined in Sec.~\ref{sec:viable}.

\subsection{Viable model consistent with the Planck result} 
\label{sec:viable}

\begin{figure}[H]
\centering
\includegraphics[width=.8\textwidth]{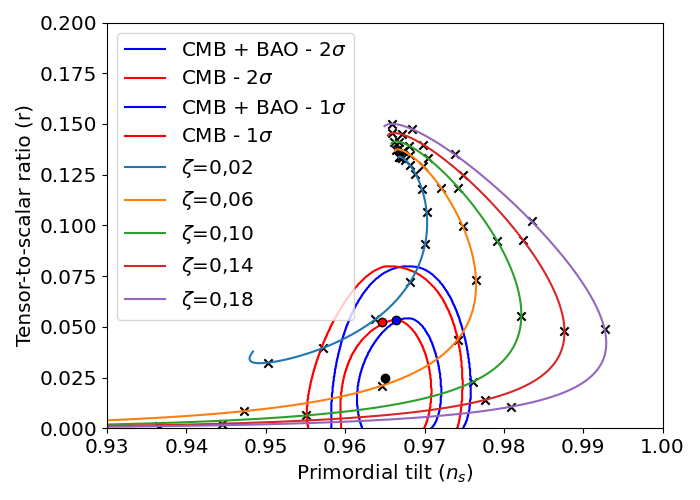}
\caption{Prediction of $n_s$ and $r$ for various values of $\zeta$ and $\lambda$. The value of $\lambda$ varies along each line from 0.04 to 0.6. The crosses mark an increase in $\lambda$ by 0.04.
Three dots represent our reference parameter points.}
\label{figPlackdata}
\end{figure}

\begin{figure}[H]
\centering
\includegraphics[width=.7\textwidth]{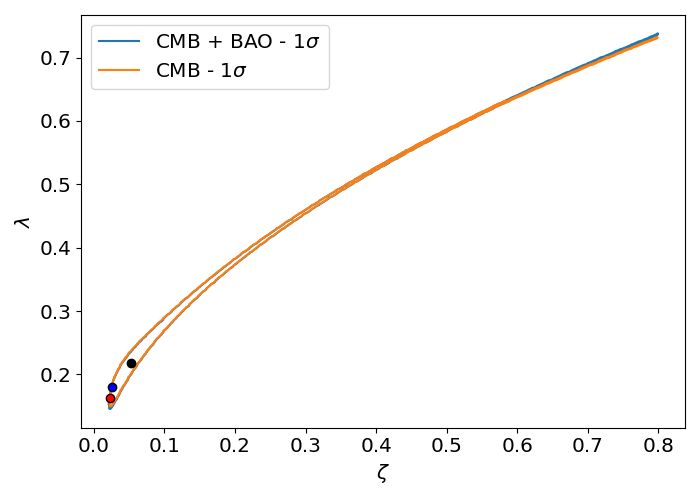}
\caption{The values of $\zeta$ and $\lambda$ consistent with the $1\sigma$ regions of Fig.~\ref{figPlackdata}. All values inside the displayed boundary are consistent.
Three dots represent our reference parameter points, which correspond to those in Fig.~\ref{figPlackdata}.}
\label{figValidParamRange}
\end{figure}

\begin{figure}[H]
\centering
\includegraphics[width=.7\textwidth]{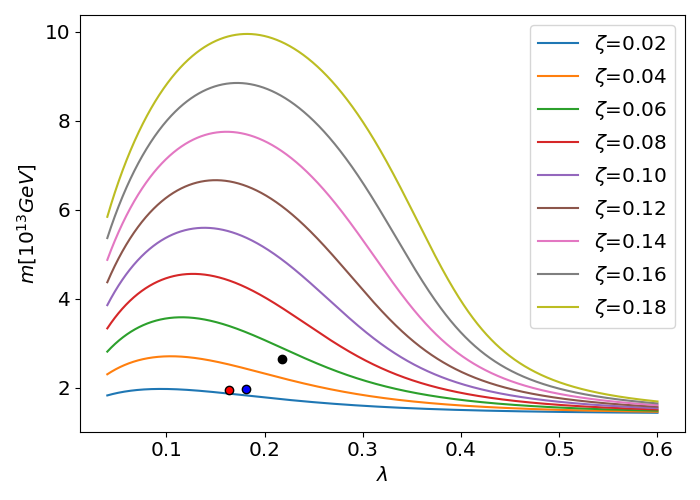}
\caption{The inflaton mass for various values of $\zeta$ and $\lambda$.
Three dots represent our reference parameter points, which correspond to those in Fig.~\ref{figPlackdata}.}
\label{figMasses}
\end{figure}

In order to compare theoretical predictions for $(n_s, r)$ with the observations, the slow-roll equation \eqref{eqSlowRollFriedmann} was integrated using the fourth order explicit Runge-Kutta scheme. We assumed that inflation lasts for 60 e-folds. The endpoint of the inflation is determined by the condition ${\rm max}\left(\epsilon\mathrm{,}\lvert\eta\rvert\mathrm{,}\lvert\alpha\rvert\mathrm{,}\lvert\beta\rvert\mathrm{,}\gamma\right)=1$. For our $\phi^2$ potential they are explicitly given by
\begin{align}
    &\epsilon = \eta = \frac{2 M_{\rm pl}^2}{\phi^2}, \\
    &\alpha = -\frac{2}{M_{\rm pl}^3}\lambda\xi_{0}m^2\phi e^{-\lambda\phi / M_{\rm pl}} = -3\lambda\zeta\frac{\phi}{M_{\rm pl}}e^{-\lambda\phi / M_{\rm pl}}, \\
    &\beta = \frac{2}{3 M_{\rm pl}^4}m^2\lambda^2\xi_{0}\phi^2 e^{-\lambda\phi / M_{\rm pl}}= \lambda^2\zeta\frac{\phi^2}{M_{\rm pl}^2}e^{-\lambda\phi / M_{\rm pl}}, \\
    &\gamma = \frac{8}{9 M_{\rm pl}^8}m^4\lambda^2\xi_{0}^2\phi^4 e^{-2\lambda\phi / M_{\rm pl}}= 2\lambda^2\zeta^2\frac{\phi^4}{M_{\rm pl}^4}e^{-2\lambda\phi / M_{\rm pl}},
\end{align}
where we have defined $\zeta\equiv \frac{2 m^2 \xi_0}{3 M_{\rm pl}^2}$.
The scalar spectral index $n_s$ and the tensor-to-scalar ratio $r$ are then evaluated through Eqs.~(\ref{ns}) and (\ref{nT}).
Furthermore, we demand that the density perturbation satisfies the condition (\ref{Ppsi}) to determine the inflaton mass $m$.
The results for $(n_s, r)$ for various values of $\lambda$ and $\zeta$ are shown in Fig.~\ref{figPlackdata}. 
We find that the results are consistent with those presented in Ref.~\cite{Jiang:2013gza}.
Red and blue lines correspond to the parameter region derived by the CMB data only and CMB + baryon acoustic oscillation (BAO) data, respectively~\cite{Planck:2018jri}.  
We see that unlike in pure GR without the Gauss-Bonnet term, the $\phi^2$ model can be made consistent with the observational bounds. 
The range of $\lambda$ and $\zeta$ consistent with the observation is presented in Fig.~\ref{figValidParamRange},
and the consistent inflaton mass is shown in Fig.~\ref{figMasses}.
This results gives us a lower bound on the inflaton mass as $m_{\rm min}\simeq 1.94\times 10^{13}\,{\rm GeV}$ for the $1\sigma$ CMB only region of Fig.~\ref{figPlackdata} and $m_{\rm min}\simeq 1.98\times 10^{13}\,{\rm GeV}$ for the $1\sigma$ joint CMB + BAO region. These points are marked in Fig.~\ref{figPlackdata}, Fig.~\ref{figValidParamRange} and Fig.~\ref{figMasses} by a red point for the CMB only region value and blue point for the joint CMB + BAO region value.

\subsection{High-frequency gravitational waves from inflaton decay}
\label{sec:high}

After inflation ends, the inflaton begins a coherent oscillation around the potential minimum $\phi=0$.
During this reheating phase, the inflaton-Gauss-Bonnet coupling induces the inflaton decay into the graviton pair~\cite{Ema:2021fdz}, which is a unique feature of our model.\footnote{
	Without the Gauss-Bonnet term, the inflaton cannot decay into the graviton pair. Instead, the inflaton annihilation to the graviton pair is a ubiquitous process~\cite{Ema:2015dka,Ema:2016hlw,Ema:2020ggo,Ghiglieri:2022rfp}, although the resulting high-frequency GW spectrum is typically much smaller than the case of inflaton decay into the graviton pair~\cite{Ema:2020ggo}.
}
Gravitons produced by the inflaton decay form stochastic GW background in the present universe.
In terms of the density parameter of the GW, $\Omega_h \equiv \rho_h/\rho_{\rm crit}$ with $\rho_h$ and $\rho_{\rm crit}$ being the GW and critical energy density, the stochastic GW background spectrum is evaluated as~\cite{Ema:2021fdz}
\begin{equation}
\label{eqGravSpectrumFormula}
	\frac{d\Omega_h}{d \ln E} = \frac{1}{\rho_{\rm crit}}\frac{16 E^4}{m^4}\frac{\Gamma\left(\phi\rightarrow2h\right)\rho_{\phi}\left(z\right)}{H\left(z\right)},
\end{equation}
where $E$ is the present GW energy, $m$ is the mass of the inflaton field, $\rho_{\phi}$ its density, $\Gamma\left(\phi\rightarrow 2h\right)$ is the decay rate of inflaton into the graviton pair and $\rho_{\rm crit} \simeq 8.0992h^2\times 10^{-47} {\rm GeV^4}$ with $h\simeq 0.67$ being the dimensionless Hubble parameter. The expression is evaluated at the redshift when the graviton pair was produced, which is given by $z=m/(2E)-1$.
Given the model parameters as determined in the previous subsection, we can calculate the stochastic GW background spectrum using \eqref{eqGravSpectrumFormula}. 
Below we explain how to evaluate the quantities appearing in the right hand side of \eqref{eqGravSpectrumFormula}.

By expanding the Gauss-Bonnet coupling function as $8\xi_0 e^{-\lambda\phi/M_{\rm pl}}\simeq 8\xi_0\left(1-\lambda\phi/M_{\rm pl}\right)$, we obtain the effective Lagrangian around the potential minimum as
\begin{align}
	\mathcal L \simeq -\frac{\phi}{\Lambda} R_{\rm GB}^2,~~~~~~\Lambda\equiv \frac{2M_{\rm pl}}{\xi_0\lambda}.
\end{align}
This term induces the perturbative inflaton decay into the graviton pair with the decay rate given by~\cite{Ema:2021fdz}
\begin{equation}
 	\Gamma\left(\phi\rightarrow2h\right)=\frac{m^7}{4\pi \Lambda^2 M_{\rm pl}^4}.
\end{equation}
On the other hand, the total decay rate of the inflaton $\Gamma_{\rm tot}$ will be parametrized in terms of the reheating temperature $T_{\rm R}$ as
\begin{equation}
\Gamma_{\rm tot}=\sqrt{\frac{\pi^2 g_* T_{\rm R}^4}{90 M_{\rm pl}^2}},
\end{equation}
where $g_*$ is the relativistic degrees of freedom.
For example, the inflaton may couple to the Standard Model gauge bosons through the coupling like $\mathcal L \sim (\phi/M)F_{\mu\nu}F^{\mu\nu}$ with some cutoff scale $M$ and this may determine the total decay rate of the inflaton, although a concrete coupling needs not be specified in the following discussion. 
Since $\Gamma_{\rm tot}$ must be higher than $\Gamma\left(\phi\rightarrow 2h\right)$ in order to avoid the constraint from the dark radiation abundance, we have a lower bound on $T_{\rm R}$. However, other than that, $T_{\rm R}$ remains to be another free parameter in the model. In order to obtain a reasonable estimate of the GW spectrum we will not solve the whole Friedmann equations during the decay period. Instead we shall approximate the Hubble parameter by 
\begin{equation}
\label{eqHubbleApproximation}
H\left(z\right)=\frac{H_{\rm ref}\left(\frac{a_{\rm ref}}{a\left(z\right)}\right)^2}{1+c\left(\frac{a_{\rm ref}}{a\left(z\right)}\right)^{1/2}\left(\frac{H_{\rm ref}}{\Gamma_{\rm tot}}\right)^{1/4}},
\end{equation}
where $c$ is a constant of order unity, which will be taken to be one for later numerical calculation. 
Here the lower index ``ref'' refers to an arbitrary reference point well after the completion of the reheating, i.e., $T\ll T_{\rm R}$. 
Eq.~(\ref{eqHubbleApproximation}) is not an exact expression but an approximate formula that extrapolates the regime $H\ll \Gamma_{\rm tot}$ and  $H\gg \Gamma_{\rm tot}$. This approximate expression is useful since everything is calculated analytically, as shown below. 
Under this approximation we can explicitly integrate the relation between time and redshift as
\begin{equation}
t\left(z\right)=\int_0^t dt^{\prime}=\int_z^{\infty}\frac{dz^{\prime}}{H\left(1+z\right)}=\frac{1}{H_{\rm ref}}\left[\frac{1}{2}\left(\frac{a}{a_{\rm ref}}\right)^2+\frac{2c}{3}\left(\frac{a}{a_{\rm ref}}\right)^{3/2}\left(\frac{H_{\rm ref}}{\Gamma_{\rm tot}}\right)^{1/4}\right].
\end{equation}
As long as we keep the reference point well below $T_{\rm R}$ we can obtain $H_{\rm ref}$ from the Friedmann equation\footnote{Reference temperature $T_{\rm ref}$ is related to the current CMB temperature $T_0$ by the entropy conservation: $g_{*s}T_{\rm ref}^3 a_{\rm ref}^3=g_{*s0}T_{0}^3a_{0}^3$. }
\begin{equation}
H_{\rm ref} = \sqrt{\frac{\pi^2 g_{*} T_{\rm ref}^4}{90 M_{\rm pl}^2}}.
\end{equation}
To evaluate the inflaton density during the decay we observe that
\begin{equation}
\rho_{\phi}\left(t\right)=\rho_{\phi}\left(\tilde{t}_{\rm ref}\right)\left(\frac{\tilde{a}_{\rm ref}}{a\left(t\right)}\right)^3 e^{-\Gamma_{\rm tot}t}=3\tilde{H}_{\rm ref}^2 M_{\rm pl}^2\left(\frac{\tilde{a}_{\rm ref}}{a\left(t\right)}\right)^3 e^{-\Gamma_{\rm tot}t},
\end{equation}
for an arbitrary another reference point well before the inflaton decay corresponding to the time $\tilde{t}_{\rm ref}$. Using \eqref{eqHubbleApproximation}\footnote{We can safely omit the 1 in the denominator because $\frac{a_{\rm ref}}{\tilde{a}_{\rm ref}}\gg 1$.} we obtain an expression explicitly independent of $\tilde{a}_{\rm ref}$:
\begin{equation}
\rho_\phi(z)=\frac{3 H_{\rm ref}^2 M_{\rm pl}^2}{c^2}\left(\frac{a_{\rm ref}}{a(z)}\right)^3\left(\frac{\Gamma_{\rm tot}}{H_{\rm ref}}\right)^{1/2}e^{-\Gamma_{\rm tot}t(z)}.
\end{equation}
Now we obtained all the ingredients to evaluate the GW spectrum (\ref{eqGravSpectrumFormula}) for given model parameters $m$, $\xi_0$ and $\lambda$. 

\begin{figure}[H]
\begin{center}
\includegraphics[width=.8\textwidth]{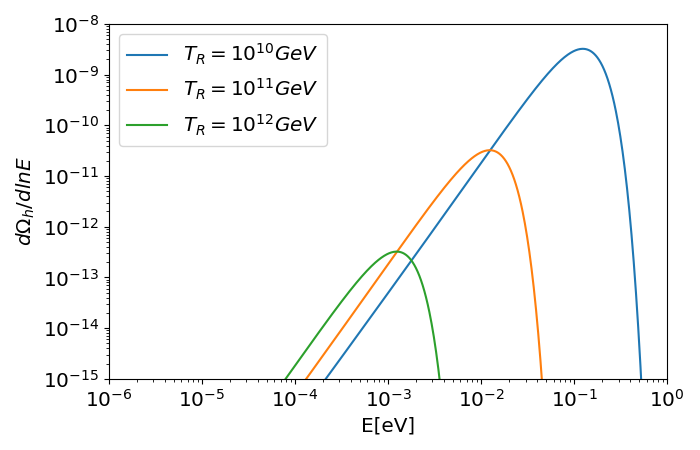}
\caption{The GW spectrum from the perturbative inflaton decay into the graviton pair in our model for $\zeta=0.053$, $\lambda=0.22$ and $T_{\rm R}= 10^{10}\,{\rm GeV}, 10^{11}\,{\rm GeV}, 10^{12}\,{\rm GeV}$ from top to bottom.}
\label{figGWspectrum}
\end{center}
\end{figure}

For numerical calculation, we choose the parameters as $\lambda=0.22$, $\zeta=0.053$ and $m=2.6\times 10^{13}\,{\rm GeV}$ which lie inside both of the observationally favored parameter regions. We marked this point by a black dot in Fig.~\ref{figPlackdata}, Fig.~\ref{figValidParamRange} and Fig.~\ref{figMasses}. For this point we have 
\begin{align}
&n_s=0.965,~~~~r=0.0247,~~~~\Gamma\left(\phi\rightarrow2h\right)=0.019\,{\rm GeV}.
\end{align}
Note that we have a lower bound on $T_{\rm R}$ as $T_{\rm R} \gg 10^8\,{\rm GeV}$, since otherwise the graviton abundance becomes too large to be consistent with the constraint on dark radiation abundance~\cite{Planck:2018vyg}.
Fig.~\ref{figGWspectrum} shows the resulting GW spectrum for $T_{\rm R}=10^{10}, 10^{11}, 10^{12}$\,GeV.
One can see that the low energy tail of the spectrum scales as $d\Omega_h/d\ln E \propto E^{5/2}$, as expected from the formula (\ref{eqGravSpectrumFormula}).
The peak energy is given by $E_{\rm peak} \sim m T_0/T_{\rm R}$ and hence it is inversely proportional to $T_{\rm R}$, while the peak spectrum is given by $\Omega_h^{\rm peak} \sim \Omega_{\rm rad} \times \Gamma(\phi\to 2h) / \Gamma_{\rm tot} \propto m^7/T_{\rm R}^2$, where $\Omega_{\rm rad}$ denotes the present radiation energy density parameter.
Thus we have $\Omega_{h}^{\rm peak} \propto E_{\rm peak}^2$ for fixed $m$ with varying $T_{\rm R}$, as seen in the figure.

\subsection{Gravitational wave background spectrum}
\label{sec:spectrum}

Now let us show the various GW background spectrum predicted in the present model, other than the one from inflaton decay to the graviton pair studied in the previous subsection.

\paragraph{Inflationary GWs}

Generically the inflation predicts the so-called primordial GWs that are generated during the de Sitter expansion  phase in very wide frequency range~\cite{Maggiore:1999vm,Maggiore:2018sht,Smith:2005mm,Boyle:2005se}. 
It extends to the present cosmological horizon scale at the low frequency limit and the comoving horizon scale of the inflation at the high frequency limit.
The overall amplitude of the spectrum is proportional to $H_{\rm inf}^2$.
GWs that enter the horizon during the radiation dominated era exhibit the flat part of the GW spectrum today.
There is a spectral break at the frequency corresponding to the comoving Hubble scale at the completion of reheating~\cite{Nakayama:2008ip,Nakayama:2008wy,Kuroyanagi:2008ye}, above which the spectrum decreases as $\Omega_h \propto f^{-2}$ with the GW frequency $f$.
This spectral break frequency is given by $\sim 2.6\times 10^2\,{\rm Hz} (T_{\rm R}/10^{10}\,{\rm GeV})$.
In the extreme high frequency limit, the spectrum is cut off at the frequency corresponding to the comoving horizon scale of the end of inflation: $f_{\rm end} \sim 10^6\,{\rm Hz} (H_{\rm end}/10^{13}\,{\rm GeV})^{1/3}(T_{\rm R}/10^{10}\,{\rm GeV})^{1/3}$.

\paragraph{Reheating GWs}

It has been shown in Refs.~\cite{Ema:2015dka,Ema:2020ggo} that the GW spectrum generically extends to higher frequency beyond $f_{\rm end}$ and that part scales as $\Omega_h \propto f^{-1/2}$.
This is purely generated by the gravitational effect: the oscillating scale factor induced by the inflaton coherent oscillation leads to the graviton particle production without any interaction beyond the Einstein-Hilbert term.
It is also interpreted as the annihilation of inflaton particles into the graviton pair.
This contribution is finally cut off at the frequency corresponding to the gravitons produced around the completion of the reheating, $f_{\rm cut} \sim 3\times 10^{13}\,{\rm Hz} (m/10^{13}\,{\rm GeV}) (10^{10}\,{\rm GeV}/T_{\rm R})$.
In the case of Gauss-Bonnet correction studied in this paper, the inflaton oscillation produces gravitons more efficiently due to the direct inflaton coupling to the gravitons through the Gauss-Bonnet term. 
In this sense, it may be regarded as another ``reheating'' source of the GWs. 
Both show nearly the same cutoff frequency, as both mechanisms produce gravitons with frequency of $m$ or $m/2$ and the production stops when the reheating is completed.

\paragraph{Bremsstrahlung GWs}

When the inflaton decays into the Standard Model particles for successful reheating, gravitons are necessarily emitted through the bremsstrahlung processes~\cite{Nakayama:2018ptw,Huang:2019lgd,Barman:2023ymn}.
The typical magnitude of the peak GW spectrum is given by $\Omega_h \sim \Omega_{\rm rad} \times m^2/(16\pi^2 M_{\rm pl}^2)$, and the peak frequency is close to the one from the inflaton decay into the gravitons.
The low frequency tail scales as $\Omega_h \propto f$.

\paragraph{GWs from thermal bath}

Particles in thermal bath scatter with each other and GWs are produced through the bremsstrahlung process associated with such scattering processes~\cite{Ghiglieri:2015nfa,Ghiglieri:2020mhm,Ringwald:2020ist}.
The dominant contribution comes from the scattering around $T\sim T_{\rm R}$ and the overall magnitude is proportional to $T_{\rm R}$.
The typical GW frequency is about 100\,{\rm GHz}, close to the CMB frequency, which is nearly independent of $T_{\rm R}$.\\

\begin{figure}
\begin{center}
   \includegraphics[width=8cm]{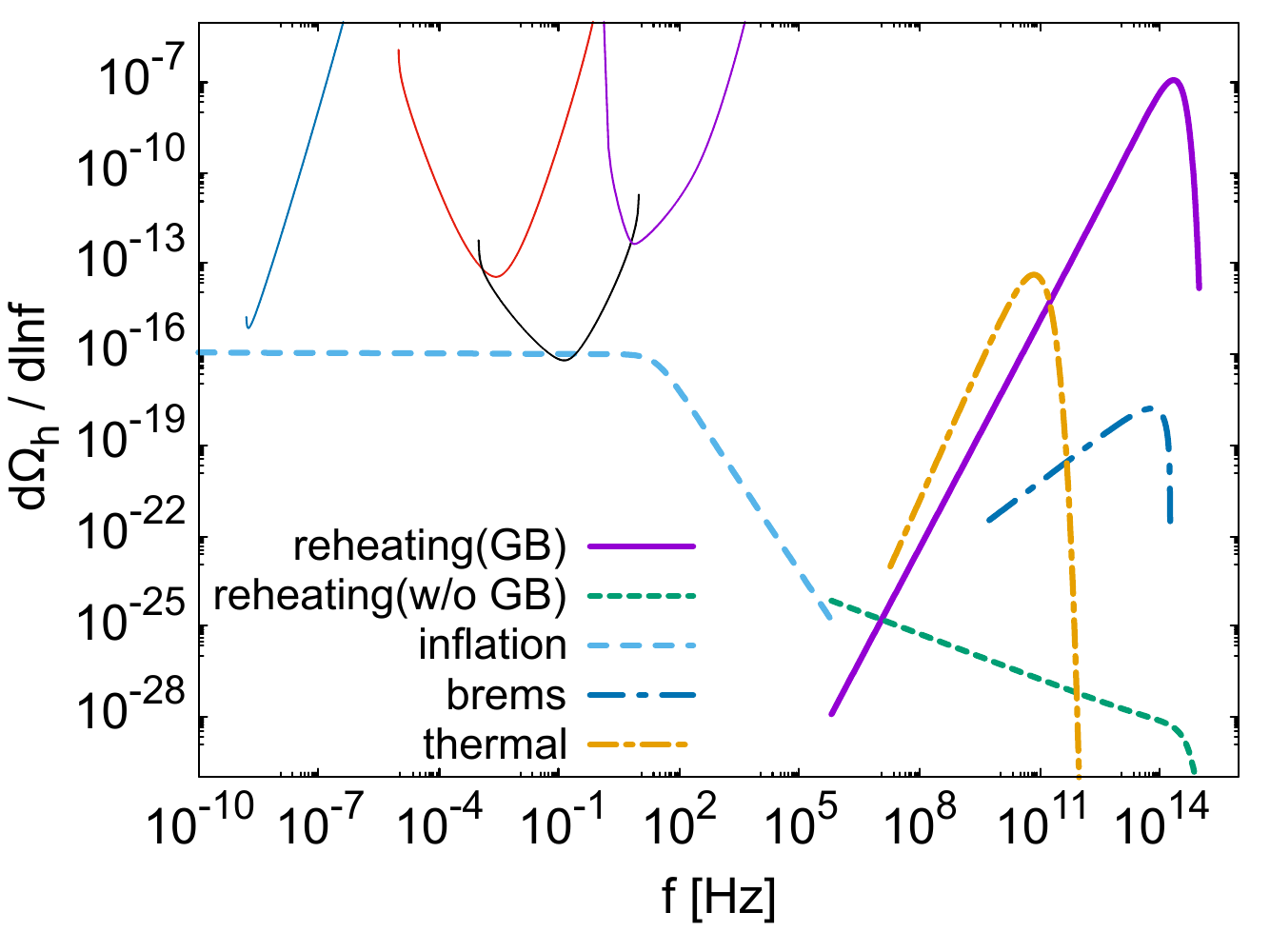}
   \includegraphics[width=8cm]{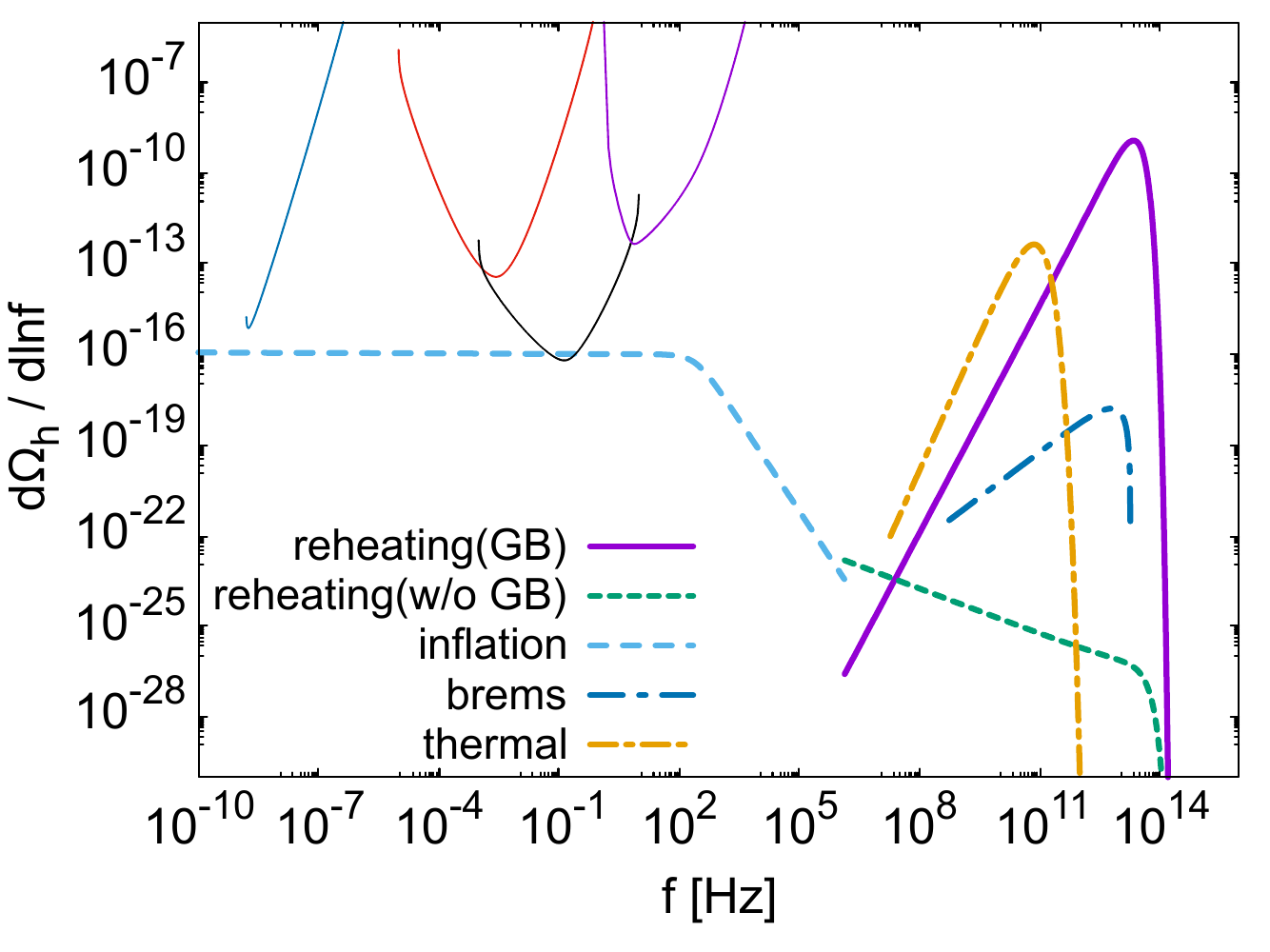}
   \includegraphics[width=8cm]{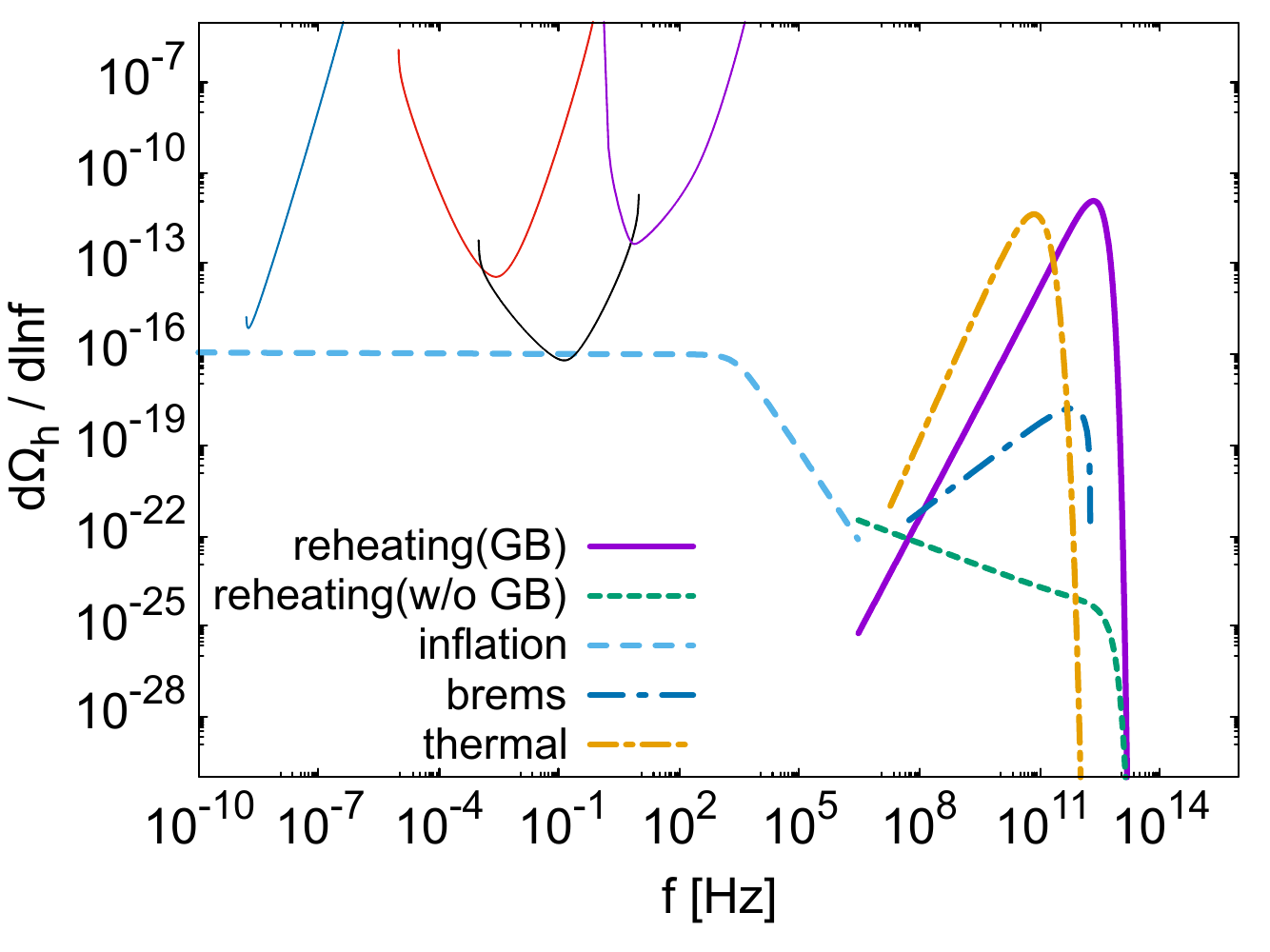}
   \includegraphics[width=8cm]{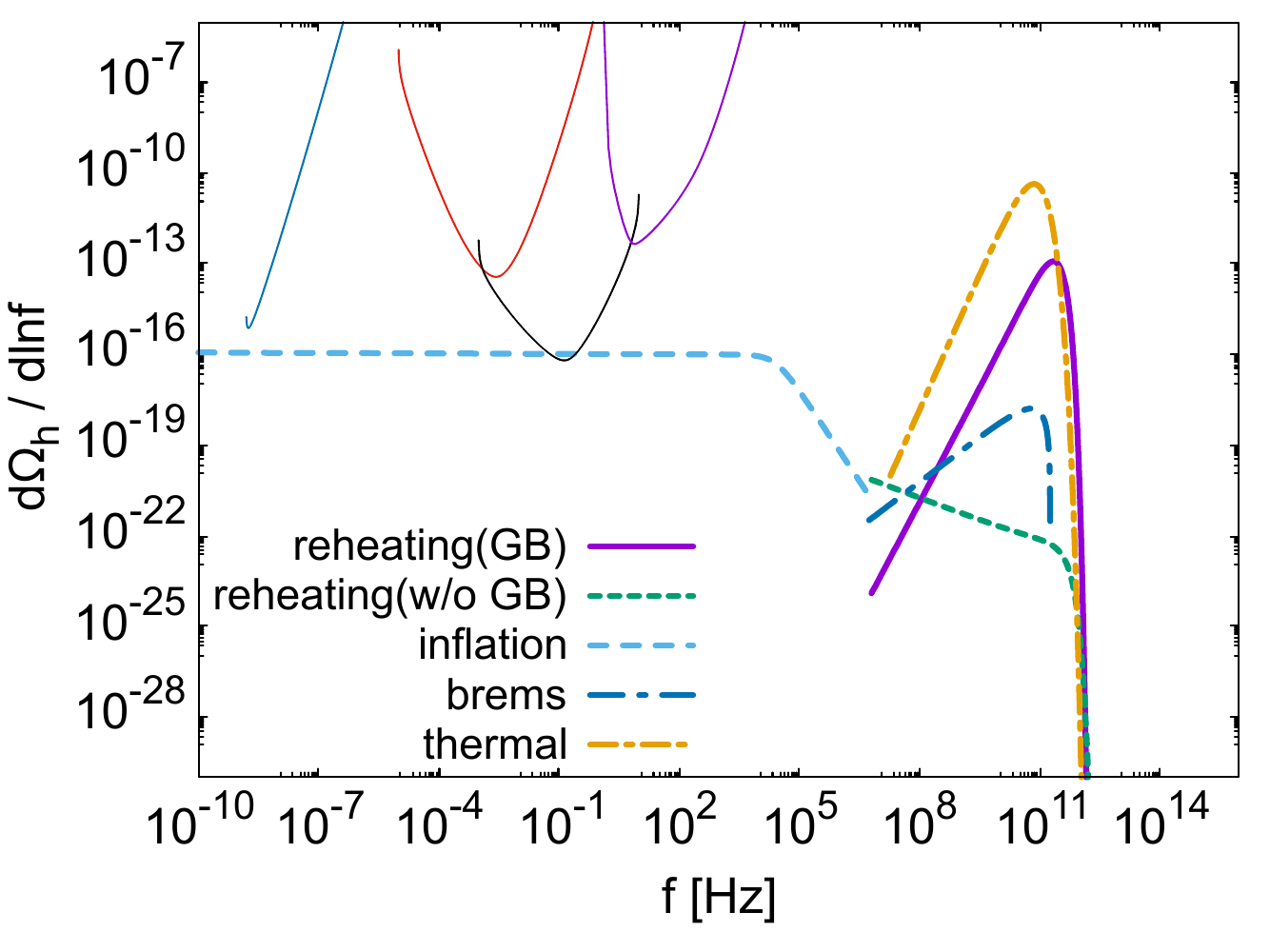}
  \end{center}
  \caption{
  GW background spectrum predicted in our model.
  We have taken $T_{\rm R}=10^9\,{\rm GeV}$ (upper left), $10^{10}\,{\rm GeV}$ (upper right), $10^{11}\,{\rm GeV}$ (lower left), $10^{12}\,{\rm GeV}$ (lower right). The new contribution from the inflaton decay to the gravitons due to the Gauss-Bonnet term is shown by the ``reheating(GB)'' line. 
The reheating GWs, which would be expected if there were no Gauss-Bonnet term, is shown by the ``reheating(w/o GB)'' line.
The inflationary GWs is shown by the ``inflation'' line, the bremsstrahlung GW is shown by the ``brems'' line and GWs from thermal bath is shown by the ``thermal'' line.
Also shown are sensitivity curves of future GW observation: SKA, LISA, DECIGO and ET from left to right.}
  \label{fig:gwb}
\end{figure}

Fig.~\ref{fig:gwb} shows these contributions compared with the GW spectrum that is a characteristic of our present model calculated in the previous subsection. 
We have taken $T_{\rm R}=10^9\,{\rm GeV}$ (upper left), $10^{10}\,{\rm GeV}$ (upper right), $10^{11}\,{\rm GeV}$ (lower left), $10^{12}\,{\rm GeV}$ (lower right). The model parameters are chosen to be the same as the previous subsection.
The new contribution from the inflaton decay to the gravitons due to the Gauss-Bonnet term is shown by the ``reheating(GB)'' line. 
The reheating GWs, which would be expected if there were no Gauss-Bonnet term, is shown by the ``reheating(w/o GB)'' line.
The inflationary GWs is shown by the ``inflation'' line, the bremsstrahlung GW is shown by the ``brems'' line and GWs from thermal bath is shown by the ``thermal'' line.
Also shown are sensitivity curves of future GW observation: SKA~\cite{Janssen:2014dka}, LISA~\cite{Bartolo:2016ami}, DECIGO~\cite{Yagi:2011wg} and ET~\cite{Punturo:2010zz} from left to right. We adopted the data in Ref.~\cite{Schmitz:2020syl,Schmitz:2020syl2} to draw these sensitivity curves.
One can see that the GWs due to the Gauss-Bonnet term is the most abundant one at the high frequency limit for $T_{\rm R}\lesssim 10^{11}\,{\rm GeV}$. For higher $T_{\rm R}$, it may be hidden by the thermal contribution.
This figure clearly shows the possibility that the high frequency GWs can be a smoking gun signature for identifying the inflation model.

\section{Conclusions}
\label{sec:conc}

We studied phenomenology of an inflation model with inclusion of the coupling between the inflaton and Gauss-Bonnet term, since such a coupling may be motivated by the quantum gravity, such as the string theory.
Such a term modifies the inflaton dynamics and change the prediction for the density perturbation, its spectral index, tensor-to-scalar ratio and so on.
We find that, with an appropriate choice of the parameters, the Gauss-Bonnet correction makes the inflation with the simple quadratic potential consistent with the CMB observation. 
Still it predicts a sizable tensor-to-scalar ratio so that it is detectable by future observations of the CMB B-mode polarization.

We pointed out that the most distinct feature of the model is imprinted in the high frequency GWs.
The inflaton-Gauss-Bonnet coupling induces the perturbative inflaton decay into the graviton pair, which results in the very abundant stochastic GWs in the present universe.
The resulting GW spectrum is shown in Fig.~\ref{fig:gwb}.
There are several different contributions to the high frequency GWs, but those produced by the inflaton decay due to the Gauss-Bonnet term is often dominant.
Interestingly, we obtain a lower bound on the reheating temperature $T_{\rm R}$ as $T_{\rm R} \gtrsim 10^8\,{\rm GeV}$ to avoid too much graviton dark radiation.
This clearly shows a potential of future observations of high frequency GWs for discriminating the inflation model. 
The high frequency GW spectrum may exhibit a rich structure depending on the inflation model and the early universe physics such as the mechanism of reheating even without violent particle production phenomena like the preheating~\cite{Kofman:1997yn}.
There are several proposals to detect high frequency GWs in experiments~\cite{Berlin:2021txa,Domcke:2022rgu,Herman:2022fau}.
Although still challenging, our findings provide motivations to improve the sensitivity for high frequency GWs.

In this paper we focused on the inflaton-Gauss-Bonnet coupling of the form $\xi(\phi) R_{\rm GB}^2$. 
Another type of the possible coupling, $\xi(\phi) R^2$, does not induce the inflaton decay into the graviton pair, while the inflaton-Chern-Simons coupling $\xi(\phi) R \widetilde R$ induces it in a similar fashion to the inflaton-Gauss-Bonnet coupling~\cite{Ema:2021fdz}.
Although the inflaton dynamics is not modified by the inflaton-Chern-Simons coupling, the cosmological tensor perturbation and also the high frequency GWs from the inflaton decay can be significantly affected.
In particular, it predicts chiral GWs for large scale tensor perturbations~\cite{Lue:1998mq,Alexander:2004wk,Satoh:2007gn}.
Our findings show that, not only superhorizon tensor fluctuations, very high frequency GWs can also be a probe of these scenarios.

\section*{Acknowledgment}

This work started during the exchange student program COLABS in Tohoku University.
This work was supported by World Premier International Research Center Initiative (WPI), MEXT, Japan.

\appendix

\section{Comparisons with other literature}  \label{sec:comparison}

There are several literature that give formulae for the scalar and tensor perturbations and their spectral indices in the Gauss-Bonnet model.
However, it is a bit inconvenient that the notations and conventions are scattered among the literature. 
Here we summarize comparisons with several references~\cite{Satoh:2008ck,Kawai:2021bye,Guo:2010jr}. 
Our main text follows the conventions of Ref.~\cite{Satoh:2008ck}. We find that all the literature give consistent results.
They are also compared with the formula for the most general second order action given in Ref.~\cite{Kobayashi:2011nu}.

\subsection*{Comparison with Ref.~\cite{Kawai:2021bye}}

Let us compare our formulae, which are based on Ref.~\cite{Satoh:2008ck}, with those given in Ref.~\cite{Kawai:2021bye}.
They defined $\sigma_i$ $(i=1,2,\dots)$, which in our notation is given by
\begin{align}
	\sigma_1 = \frac{H \dot \xi}{M_{\rm pl}^2} = \sigma,~~~~~~\sigma_2 = \frac{\dot\sigma}{H\sigma} = \frac{\ddot\xi}{\sigma M_{\rm pl}^2} - \epsilon_H,
\end{align}
where $\epsilon_H = -\dot H/H^2$.
First note that our $A_\psi^2$ (\ref{Apsi}) is rewritten as
\begin{align}
	A_\psi^2 = \frac{6a^2 X^2}{Y^2}\left( \frac{\rho^2}{6} + \frac{\sigma^2}{16X} \right).
\end{align}
By noting that $\rho^2$ is rewritten as
\begin{align}
	\rho^2 = \frac{\dot\phi^2}{H^2 M_{\rm pl}^2} = 2\epsilon_H - \frac{\sigma}{2} + \frac{\sigma\sigma_2}{2} - \frac{\sigma\epsilon_H}{2},
\end{align}
we find that our $A_\psi^2$ is equal to $A_\zeta^2$ given in Eq.~(A14) of Ref.~\cite{Kawai:2021bye}. Similarly, $C_\psi^2$ (\ref{Cpsi}) is rewritten as
\begin{align}
	C_\psi^2 &= \frac{2a^2}{A_\psi^2}\left[ (1+\epsilon_H)\frac{X^2}{Y}-\sigma\sigma_2\frac{X}{Y}\left(1-\frac{3X}{4Y}\right)-1+\frac{\sigma(\sigma_2+\epsilon_H)}{2} \right],\\
	&= 1-\frac{a^2\sigma^2}{2A_\psi^2 Y^2}\left[\epsilon_H\left( 1-\frac{5\sigma}{8}\right) + \frac{\sigma}{8}(1-\sigma_2) \right].  \label{Czeta}
\end{align}
This is equal to $C_\zeta^2$ given in Eq.~(A15) of Ref.~\cite{Kawai:2021bye}. 
It is not difficult to check that our $A_T^2$ (\ref{AT}) and $C_T^2$ (\ref{CT}) are equal to $A_t^2$ and $C_t^2$ given in Eq.~(A20) of Ref.~\cite{Kawai:2021bye}.

\subsection*{Comparison with Ref.~\cite{Guo:2010jr}}

Let us compare our formulae, which are based on Ref.~\cite{Satoh:2008ck}, with those given in Ref.~\cite{Guo:2010jr}.
Their $z_{\mathcal R}^2$ [Eq.(19) of Ref.~\cite{Guo:2010jr}] should be the same as our $A_\psi^2$. By noting that their $\xi$ corresponds to our $\xi/8$, their $z_R^2$ is calculated as
\begin{align}
	z_{\mathcal R}^2 = a^2 \frac{\dot\phi^2 + 6\Delta \dot\xi H^3}{(1-\Delta/2)^2 H^2}
	=\frac{6a^2 X^2}{Y^2}\left( \frac{\rho^2}{6} + \frac{\sigma^2}{16X} \right),
\end{align}
where $\Delta\equiv \dot\xi H / (2-\dot\xi H) = \sigma/(2-\sigma)$. This is shown to be equal to $A_\psi^2$ (\ref{Apsi}).
Also their $c_{\mathcal R}^2$ [Eq.(19) of Ref.~\cite{Guo:2010jr}] should be the same as our $C_\psi^2$. It is calculated as
\begin{align}
	c_{\mathcal R}^2 &= 1 + \frac{8\Delta\dot\xi H \dot H +2\Delta^2 H^2(\ddot\xi-H\dot\xi)}{\dot\phi^2 + 6\Delta\dot\xi H^3}\\
	&=1-\frac{a^2\sigma^2}{2A_\psi^2 \left(1-\frac{3\sigma}{4}\right)^2}\left[\epsilon_H\left(1-\frac{5\sigma}{8}\right) + \frac{\sigma}{8}(1-\sigma_2)\right]
\end{align}
It is equal to (\ref{Czeta}) and hence to our $C_\psi^2$.
It is not difficult to check that our $A_T^2$ (\ref{AT}) and $C_T^2$ (\ref{CT}) are equal to $z_T^2$ and $c_T^2$ given in Eqs.~(25) and (26) of Ref.~\cite{Guo:2010jr}.

Under the slow-roll approximation, Ref.~\cite{Guo:2010jr} also gives the scalar spectral index as
\begin{align}
	n_{\mathcal R} = 1-2\epsilon_1 - \frac{2\epsilon_1\epsilon_2-\delta_1\delta_2}{2\epsilon_1-\delta_1},
	\label{nR_GS}
\end{align}
where
\begin{align}
	\epsilon_1 = \epsilon_H = -\frac{\dot H}{H^2},~~~~~~\epsilon_2=\frac{\dot\epsilon_1}{H\epsilon_1},
	~~~~~~\delta_1=\frac{\dot\xi H}{2 M_{\rm pl}^2} = \frac{\sigma}{2},~~~~~~\delta_2=\frac{\dot\delta_1}{H\delta_1}.
\end{align}
In terms of our parametrization, they are given by
\begin{align}
	&\epsilon_1 = \epsilon + \frac{\alpha}{3},\\
	&\delta_1 = \frac{\sigma}{2} = -\frac{2}{3}\alpha-\frac{1}{2}\gamma,\\
	&\epsilon_2 = \frac{1}{\epsilon_1}\left( 4\epsilon^2-2\epsilon\eta -\eta\alpha +\frac{4}{3}\epsilon\alpha - \epsilon\beta - \frac{1}{4}\eta\gamma - \frac{1}{3}\alpha\beta \right),\\
	&\delta_2 = -\eta - \frac{4\alpha}{3}-\beta+\frac{2\beta}{\sigma}\left(2 \epsilon + \frac{4\alpha}{3} + \frac{\gamma}{2} \right).
\end{align}
By substituting them into (\ref{nR_GS}), we obtain
\begin{align}
	n_{\mathcal R} = 1 -6\epsilon + 2\eta + \frac{2\alpha}{3} + 2\beta.
\end{align}
It coincides with our $n_s$ (\ref{ns}). The tensor spectral index is given by
\begin{align}
	n_T = -2\epsilon_1 = -2\epsilon - \frac{2\alpha}{3},
\end{align}
which coincides with our $n_T$ (\ref{nT}).

\subsection*{Comparison with Ref.~\cite{Kobayashi:2011nu}}

Ref.~\cite{Kobayashi:2011nu} introduced the most general inflation models with second-order field equations.
The inflation with the Gauss-Bonnet correction falls into this category.
In the notation of Ref.~\cite{Kobayashi:2011nu}, noting that their $\xi$ corresponds to our $-\xi/16$, our model is given by
\begin{align}
	&K = X_\phi -\frac{1}{2}\xi^{(4)} X_\phi^2(3-\ln X_\phi), \\
	&G_3 = -\frac{1}{4}\xi^{(3)} X_\phi(7-3\ln X_\phi),\\
	&G_4 = \frac{M_{\rm pl}^2}{2}  -\frac{1}{4}\xi^{(2)} X_\phi(2-\ln X_\phi),\\
	&G_5 = \frac{1}{4}\xi^{(1)} \ln X_\phi,
\end{align}
where $X_\phi \equiv -(\partial\phi)^2/2$ and $\xi^{(n)}$ denotes the $n$-th derivative of $\xi(\phi)$ with respect to $\phi$.

For the tensor perturbation, $\mathcal G_T$ and $\mathcal F_T$ in Ref.~\cite{Kobayashi:2011nu} correspond to our $A_T^2/a^2$ and $A_T^2 C_T^2/a^2$ up to the $M_{\rm pl}^2$ factor, respectively. By direct calculation with Eqs.~(4.4) and (4.5) of Ref.~\cite{Kobayashi:2011nu}, we find
\begin{align}
	\mathcal G_T = M_{\rm pl}^2\left(1 - \frac{\sigma}{2}\right),~~~~~~\mathcal F_T = M_{\rm pl}^2\left(1 - \frac{\ddot\xi}{2M_{\rm pl}^2}\right),
\end{align}
which actually coincide with our $A_T^2/a^2$ and $A_T^2 C_T^2/a^2$.

For the scalar perturbation, $\mathcal G_S$ and $\mathcal F_S$ in Ref.~\cite{Kobayashi:2011nu} correspond to our $A_\psi^2/a^2$ and $A_\psi^2 C_\psi^2/a^2$ up to the $M_{\rm pl}^2$ factor, respectively. To calculate $\mathcal G_S$ and $\mathcal F_S$ through Eqs.~(4.32) and (4.33) of Ref.~\cite{Kobayashi:2011nu}, we first need $\Sigma$ and $\Theta$, which were defined in Eqs.~(4.25) and (4.26) in Ref.~\cite{Kobayashi:2011nu}. They are calculated as
\begin{align}
	&\Sigma = 3H^3 \dot\xi - 3H^2 M_{\rm pl}^2 + \frac{\dot\phi^2}{2} = -3H^2M_{\rm pl}^2 \left(1-\sigma-\frac{\rho^2}{6}\right), \\
	&\Theta = HM_{\rm pl}^2\left(1-\frac{3\sigma}{4}\right).
\end{align}
After direct calculation, we find
\begin{align}
	&\mathcal G_S = 3 M_{\rm pl}^2\left(1-\frac{\sigma}{2}\right)\left[ 1- \frac{ \left(1-\sigma-\frac{\rho^2}{6}\right)\left(1-\frac{\sigma}{2}\right)}{ \left(1-\frac{3\sigma}{4}\right)^2} \right],\\
	&\mathcal F_S =  M_{\rm pl}^2\left[ \frac{1}{a^2}\frac{d}{d\tau}\left(\frac{a\left(1-\frac{\sigma}{2}\right)^2}{H\left(1-\frac{3\sigma}{4}\right)}\right)- \left(1-\frac{\ddot\xi}{2M_{\rm pl}^2} \right) \right],
\end{align}
which actually coincide with our $A_\psi^2/a^2$ and $A_\psi^2 C_\psi^2/a^2$.



\end{document}